\documentstyle[12pt]{article}
\begin{document}
\def\thefootnote{\fnsymbol{footnote}}
\begin{flushright}
KANAZAWA-02-37  \\ 
December, 2002
\end{flushright}
\vspace*{2cm}
\begin{center}
{\LARGE\bf  Soft SUSY breaking masses 
in a unified model with doublet-triplet splitting}\\
\vspace{1 cm}
{\Large Daijiro Suematsu}
\footnote[1]{e-mail: suematsu@hep.s.kanazawa-u.ac.jp}
\vspace*{1cm}\\
{\it Institute for Theoretical Physics, Kanazawa University,\\
        Kanazawa 920-1192, Japan}\\    
\end{center}
\vspace{1cm}
{\Large\bf Abstract}\\  
We study soft supersymmetry breaking parameters in a supersymmetric 
unified model which potentially solves the doublet-triplet splitting 
problem. In the model the doublet-triplet splitting is solved by the
discrete symmetry which is allowed to be introduced due to the direct
product structure of the gauge group.
The messenger fields for the gauge mediated supersymmetry breaking 
are naturally embedded in the model.  
The discrete symmetry required by the doublet-triplet splitting 
makes the gaugino masses non-universal and also induces a different mass 
spectrum for the scalar masses from the ordinary minimal gauge mediation 
model. Independent physical $CP$ phases can remain in the 
gaugino sector even after the $R$-transformation. 
\newpage
\setcounter{footnote}{0}
\def\thefootnote{\arabic{footnote}}
\section{Introduction} 
Supersymmetry is now considered to be the most promising candidate
for the solution of the gauge hierarchy problem. Although we have no
direct evidence of the supersymmetry still now, the unification shown by the 
gauge couplings in the minimal supersymmetric standard model (MSSM)
may be considered as its indirect signal.
When we consider grand unified models such as SU(5), SO(10) {\it etc.} 
based on this gauge coupling unification,
we are often annoyed by the doublet-triplet splitting problem \cite{dt}.  
The reason is that, as stressed in \cite{w}, we cannot easily introduce 
a suitable symmetry to resolve the doublet-triplet degeneracy 
in a consistent way with the unified gauge structure.
Recently, it has been pointed out that the doublet-triplet splitting
problem can be solved by extending the gauge structure such as 
a deconstruction model \cite{w} or introducing extra dimensions \cite{kaf}.
Since the doublet-triplet splitting problem is almost general in the grand 
unified models including the superstring model, it seems to be 
interesting to find models which can solve this problem and 
also to investigate the phenomenological features in such models.
 
In this article we propose a supersymmetry breaking scenario 
which is naturally
introduced in a unified model which can solve the doublet-triplet
splitting problem. 
The model is constructed by extending the deconstruction model 
given in \cite{w}.
The similar structure to the minimal gauge mediated supersymmetry 
breaking model \cite{mgm1,mgm2,exmgm1,exmgm2} is
automatically built in the model.
In our model gaugino masses seem to be generally non-universal 
and then have non-universal phases due to the gauge structure which 
is required to realize the doublet-triplet splitting. 
The soft scalar masses can also 
have a different spectrum from the ordinary one keeping the flavor blindness.
We study the general feature of the supersymmetry breaking parameters
in addition to the structure of the $CP$ phases in this model.

This paper is organized as follows. In section 2 we define our model and 
explain how the doublet-triplet splitting can be realized. 
In section 3 we derive the soft supersymmetry breaking parameters based 
on such a feature of the model and give some comments on 
their phenomenological aspects. Section 4 is devoted to the summary.

\section{A SUSY model with the doublet-triplet splitting}
We consider a model with a direct product gauge structure such as 
${\cal G}=$SU(5)$^\prime\times$SU(5)$^{\prime\prime}$ and a
global discrete symmetry $F$ which commutes with this gauge symmetry \cite{w}.
Under this gauge structure we introduce bifundamental chiral superfields 
$\Phi_1(\bar{\bf 5},{\bf 5})$ and $\Phi_2 ({\bf 5}, \bar{\bf 5})$,
an adjoint Higgs chiral superfield $\Sigma({\bf 1}, {\bf 24})$, 
three sets of chiral superfields 
$\Psi_{10}({\bf 10}, {\bf 1})+\Psi_{\bar 5}(\bar{\bf 5}, {\bf 1})$ 
which correspond to three generations of
quarks and leptons, a set of chiral superfield 
$H({\bf 5}, {\bf 1})+\tilde H({\bf 1}, \bar{\bf 5})$ which contains 
Higgs doublets, and also a set of chiral superfield 
$\bar\chi(\bar{\bf 5}, {\bf 1})+\chi({\bf 1}, {\bf 5})$
in order to cancel the gauge anomaly induced by the above contents. 
We additionally introduce several singlet chiral superfields
$S_\alpha$. These are summarized in Table 1.
In order to induce the symmetry breaking at the high energy scale 
we introduce a superpotential such as
\begin{equation}
W_1= M_\phi{\rm Tr}\left(\Phi_1\Phi_2\right) 
+{1\over 2}M_\sigma{\rm Tr}\left(\Sigma^2\right)
+\lambda~{\rm Tr}\left(\Phi_1\Sigma\Phi_2 +{1\over 3}\Sigma^3\right) 
\end{equation}
The scalar potential based on this $W_1$ can be easily obtained as
\begin{eqnarray}
&&V={\rm Tr}~\vert M_\phi\phi_1+\lambda\phi_1\sigma + y\vert^2
+ {\rm Tr}~\vert M_\phi\phi_2+\lambda\sigma\phi_2 + x \vert^2 \nonumber \\
&&\hspace*{0.5cm} +{\rm Tr}~\vert M_\sigma\sigma 
+\lambda\phi_1\phi_2+\sigma^2 + z \vert^2, 
\end{eqnarray}
where $\phi_{1,2}$ and $\sigma$ are the scalar components of
$\Phi_{1,2}$ and $\Sigma$, respectively. They are traceless and 
$x,~y$ and $z$ are the Lagrange multipliers for these traceless
conditions.

We try to find a non-trivial and physically interesting solution of 
the minimum of this scalar potential.
The conditions for it can be written in such a way as
\begin{eqnarray}
&&\phi_2={x\over y}\phi_1, \label{eqa01}\\ 
&&M_\phi\phi_1+\lambda\phi_1\sigma+ y=0, \label{eqa02}\\ 
&&M_\sigma\sigma +\lambda\left(\sigma^2+{x\over y}\phi_1^2\right)+z=0, 
\label{eqa03}
\end{eqnarray}
where the Lagrange multipliers $y$ and $z$ are determined as
\begin{equation}
y=-{\lambda\over 5}~{\rm Tr}\left(\phi_1\sigma\right), \quad
z=-{\lambda\over 5}~{\rm Tr}\left(\sigma^2
-{5x\over \lambda{\rm Tr}\left(\phi_1\sigma\right)}\phi_1^2\right),
\end{equation}
where $x$ remains as a free parameter. We restrict ourselves to consider 
a special direction in the field space such that $\phi_1=\kappa\sigma$ 
and we also assume that $M_\sigma=M_\phi(1+x\kappa^2/y)$ is satisfied.
Then, along this direction eqs.~(\ref{eqa02}) and (\ref{eqa03}) 
become consistent with each other and they are
reduced to an interesting equation such as
\begin{equation}
M_\phi\sigma + \lambda\sigma^2 
-{\lambda\over 5}~{\rm Tr}\left(\sigma^2\right)=0. 
\end{equation}
This equation has the same form as the potential minimum condition 
for the adjoint Higgs scalar in the ordinary supersymmetric SU(5). 
It is well-known that there are three supersymmetric degenerate 
independent solutions in this equation.
The most interesting one can be written as 
\begin{equation}
\sigma=\tilde M~{\rm diag}~(2,~2,~2,~-3,~-3),
\label{eqa}
\end{equation}
where $\tilde M$ is defined as $\tilde M=M_\phi/\lambda$.
In the present discussion we adopt this solution. Using this $\sigma$,
other fields can be determined as
\begin{equation}
\phi_1=\kappa\sigma, \qquad 
\phi_2={1\over \kappa}\left({M_\sigma\over M_\phi}-1\right)\sigma.
\label{eqa0}
\end{equation}
We have an unfixed parameter $\kappa$ in this solution.
However, if we assume that this model is obtained as a result of a suitable
deconstruction, $\kappa$ can be determined as discussed below. 
There is no $D$-term contribution to $V$ from these vacuum expectation
values (VEVs) in eqs.~(\ref{eqa}) and (\ref{eqa0}) and then the
supersymmetry is conserved at this stage. 

It is convenient to use the deconstruction method in order to see what kind of 
discrete symmetry remains unbroken when these VEVs are induced \cite{w}.
We consider the theory space represented by the moose diagram which is
composed of the $n$ sites $Q_i$ placed on the vertices of an $n$-polygon and 
one site on its center $P$ of this polygon. 
We assign SU(5)$^\prime$ on the site $P$ and 
SU(5)$^{\prime\prime}$ on each site $Q_i$ and also put a bifundamental 
chiral superfield $\Phi_i$ on each link from $P$ to $Q_i$. 
On each link from $Q_i$ to $Q_{i+1}$ we put the adjoint Higgs chiral superfield
$\Sigma$ of SU(5)$^{\prime\prime}$. For the later discussion, we 
may consider the unitary link variables $U_i\equiv \exp({i\phi_i/\tilde M})$ 
and ${\cal W}\equiv \exp({-i\sigma/\tilde M})$.  
Here we introduce an equivalence relation
only for the boundary points of the polygon by the $2\pi/n$ rotation and 
we identify this $Z_n$ symmetry with the above mentioned discrete
symmetry $F$.
The equivalence relation defined by $F$ makes $\Sigma$
independent of $i$ or, equivalently, invariant under $F$.
This makes us consider the reduced theory space composed of only three
sites $P$, $Q_1$ and $Q_2$, in which the field contents become equivalent 
to the one given in Table 1.
If we use ${\cal W}$ introduced above, this equivalence relation requires that 
 ${\cal W}^n=1$ is satisfied.
Thus we can write ${\cal W}$ as
\begin{equation}
{\cal W}={\rm diag}~(e^{2i\rho},~e^{2i\rho},~e^{2i\rho},
~e^{-3i\rho},~e^{-3i\rho}),
\end{equation} 
where $e^{i\rho}$ is the $n$-th root of unity.
If we assume that our model is obtained as a result of the above discussed
deconstruction, the condition $U_i {\cal W}U_{i+1}^{-1}=1$ should 
be satisfied for $i=1$, which means that the holonomy around 
each two-dimensional 
plaquette is equal to 1.\footnote{This corresponds to the energy minimum
condition from the viewpoint of the lattice gauge \cite{w}.} 
This requirement is interpreted in our vacuum defined by
eqs.~(\ref{eqa}) and (\ref{eqa0}) as an additional condition 
$-\phi_1+\sigma+\phi_2=0$, which can be transformed into a condition for the
$\kappa$ in $\kappa^2-\kappa+1-M_\sigma/M_\phi=0$.  
   
Now we consider the transformation property of this vacuum under the
gauge transformation such as
\begin{equation}
U_i^\prime=\omega^\prime~U_i ~(\omega^{\prime\prime})^{-1},  \qquad
{\cal W}^\prime=\omega^{\prime\prime}~{\cal W}~ (\omega^{\prime\prime})^{-1},
\end{equation}  
where $\omega^\prime$ and $\omega^{\prime\prime}$ are the group elements of
SU(5)$^\prime$ and SU(5)$^{\prime\prime}$, respectively.
The invariance of $U_i$ and ${\cal W}$ shows that the group elements
$\omega$ of the unbroken gauge group satisfy the condition: 
$\omega=\omega^\prime=\omega^{\prime\prime}$
and $[\omega, ~{\cal W}]=0$. Since we take the VEVs of Higgs scalar fields 
in such a way as eqs.~(\ref{eqa}) and (\ref{eqa0}), 
the unbroken gauge group is 
${\cal H}$=SU(3)$\times$SU(2)$\times$U(1) which is a
subgroup of the diagonal sum SU(5) of ${\cal G}$.
Next we consider a discrete symmetry $F^\prime$ as a diagonal subgroup of 
$F\times G_{{\rm U(1)}^{\prime\prime}}$ where
$G_{{\rm U(1)}^{\prime\prime}}$ is a discrete subgroup of a hypercharge
in SU(5)$^{\prime\prime}$.
If we write the group elements of $F$ and $G_{{\rm
U(1)}^{\prime\prime}}$ as $f$ and $\omega_D$, the transformation of $U_i$ 
due to $F^\prime$ can be written as
\begin{equation}
U_i^\prime=(f U_i)~\omega_D^{-1}=U_{i+1}\omega_D^{-1}.
\end{equation}
If we take $\omega_D$ as ${\cal W}$, we find that $U_i$ is invariant
under this transformation due to the relation $U_i {\cal W}U_{i+1}^{-1}=1$ 
and $F^\prime$ remains unbroken. 
The invariance of ${\cal W}$ is also clear.
Thus we can conclude that in this model the symmetry ${\cal G}\times F$ 
breaks down into ${\cal H}\times F^\prime$ by considering the vacuum 
defined by eqs.~(\ref{eqa}) and (\ref{eqa0}).
Since the definition of $F^\prime$ contains the disctrete subgroup of 
U(1)$^{\prime\prime}$ in SU(5)$^{\prime\prime}$ as its component,
every field which has a nontrivial transformation property with respect 
to SU(5)$^{\prime\prime}$ can have different charges. 
We assign the charges of $F^\prime$ for every field as shown in Table 1.
\begin{figure}[tb]
\begin{center}
\begin{tabular}{|l|c|c|c|c|}\hline
         & ${\cal F}({\cal G}~{\rm rep.})$  &  F    
& \multicolumn{2}{|c|}{$F^\prime$} \\\cline{4-5}
   &&& ${\bf 3}\in {\bf 5}$ or $\bar{\bf 3}\in\bar{\bf 5}$ & 
${\bf 2}\in{\bf 5}$ or $\bar{\bf 2}\in\bar{\bf 5}$ \\\hline
{\rm Quarks/Leptons}& $\Psi_{10}^j({\bf 10}, {\bf 1})$ &  $\alpha$ & 
$\alpha$ & $\alpha$ \\
$(j=1\sim 3)$   & $\Psi_{\bar 5}^j(\bar{\bf 5}, {\bf 1})$ &$\beta$ & $\beta$ 
& $\beta$ \\\hline 
{\rm Higgs fields}& $H({\bf 5}, {\bf 1})$ & $\gamma$ & $\gamma$ & $\gamma$\\
           & $\tilde H({\bf 1}, \bar{\bf 5})$ & $\xi$ 
& $\xi+2a$ & $\xi-3a$ \\ \hline
{\rm Messenger fields}& $\bar\chi(\bar{\bf 5}, {\bf 1})$ & $\delta$ & 
$\delta$ & $\delta$\\ 
              & $\chi({\bf 1}, {\bf 5})$ & $\zeta$ & 
$\zeta+2b$ & $\zeta-3b$
 \\ \hline
{\rm Bifundamental field}  & $\Phi_1(\bar{\bf 5},{\bf 5})$ & $\eta$ 
& $\eta+2c$ &  
$\eta-3c$ \\
            & $\Phi_2({\bf 5},\bar{\bf 5})$ & $\sigma$ & $\sigma+2d$ & 
$\sigma-3d$ \\\hline
{\rm Adjoint Higgs field}  & $\Sigma({\bf 1},{\bf 24})$ & 0 & 
\multicolumn{2}{|c|}{0}     \\\hline
{\rm Singlets}   & $S_1({\bf 1}, {\bf 1})$& $e$ &  \multicolumn{2}{|c|}{$e$}\\
                 &$S_2({\bf 1}, {\bf 1})$ & $f$ & \multicolumn{2}{|c|}{$f$}\\
\hline
\end{tabular}
\vspace*{3mm}\\
{\footnotesize Table 1~ Discrete charge assignment for the chiral 
superfields.} 
\end{center}
\end{figure} 

In order to solve the doublet-triplet splitting problem, 
only the color triplet Higgs chiral superfields $H_3$ 
and $\tilde H_{3}$
except for the ordinary Higgs chiral superfields $H_2$ and $\tilde H_{2}$ 
should become massive when the above discussed symmetry breaking occurs.
We should also require the conditions on 
$F^\prime$ to satisfy various 
phenomenological constraints in a consistent way with this realization.
We impose the following conditions.\\
(i)~Each term in the superpotential $W_1$ should exist and this
requirement imposes the conditions:
\begin{equation}
\eta+\sigma +2(c+d)=0, \qquad \eta+\sigma -3(c+d)=0.
\end{equation}
(ii)~The gauge invariant bare mass terms of the fields such as 
$\Psi_{\bar 5}H$, $H\bar\chi$, $\tilde H\chi$ and $S_\alpha S_\beta$
should be forbidden. These conditions are summarized as,
\begin{eqnarray}
&&\beta+\gamma\not=0, \qquad \gamma+\delta\not=0,\qquad  
\xi+\zeta+2(a+b)\not=0, \nonumber \\
&&\xi+\zeta-3(a+b)\not=0, \qquad 2e\not=0,\qquad 2f\not=0, \qquad e+f\not=0.
\end{eqnarray}
(iii)~To realize the doublet-triplet splitting, Yukawa coupling 
$\Phi_1 H_2\tilde H_{2}$ should be forbidden although
$\Phi_1H_3\tilde H_{3}$ is allowed. This gives the conditions such as
\begin{equation}
\gamma+\xi-3a+\eta-3c\not=0,\qquad \gamma+\xi+2a+\eta+2c= 0.
\end{equation}
(iv)~Yukawa couplings of quarks and leptons, that is, 
$\Psi_{10}\Psi_{10}H_2$ and $\Psi_{10}\Psi_{\bar 5}\tilde H_{\bar
2}\Phi_1$ should exist. This requires
\begin{equation}
 2\alpha + \gamma=0, \qquad 
\alpha+\beta+\xi-3a+\eta-3c=0.
\end{equation} 
(v)~The fields $\chi$ and $\bar\chi$ should be massless
at the ${\cal G}$ breaking scale and play the role of the messenger
fields of the supersymmetry breaking which is assumed to occur in the
$S_\alpha$ sector.
These require both the absence of $\Phi_2\chi\bar\chi$ and the existence 
of the coupling $\Phi_2S_\alpha\chi\bar\chi$. These conditions
 can be written as
\begin{eqnarray}
&&\delta+\zeta+2b+\sigma+2d\not=0,\qquad \delta+\zeta-3b+\sigma-3d\not=0,
\nonumber \\
&&\delta+\zeta+2b+\sigma+2d+e=0,\qquad \delta+\zeta-3b+\sigma-3d+f=0.
\label{mess}
\end{eqnarray}
(vi)~The neutrino should be massive and the proton should be stable.
This means that $\Phi_{\bar 5}^2H_2^2$ should exist and
$\Psi_{10}\Psi_{\bar 5}^2$ and $\Psi_{10}^3\Psi_{\bar 5}$ should be
forbidden \cite{w}. These require 
\begin{equation}
2(\beta+\gamma)=0, \qquad \alpha+ 2\beta\not=0, \qquad
3\alpha+ \beta\not= 0.
\end{equation}
Every equation should be understood up to modulus $n$ when we take
$F^\prime=Z_n$.

We can easily find an example of the consistent solution 
for these constraints. 
For example, if we take $F^\prime=Z_{20}$, such an example can be
given as
\begin{eqnarray}
&&\alpha=\delta=\eta=b=-e=1,\qquad \sigma=\xi=\zeta=-a=3, \nonumber \\
&&\gamma=-c=-2, \qquad d=-f=6, \qquad \beta=-8, 
\end{eqnarray}
where these charges should be understood up to the modulus $20$.
We have not taken account of the anomaly of $F^\prime$.
Although this anomaly cancellation seems to require the introduction of
new fields and impose the additional constraints on the charges, 
it does not affect the result of the present phenomenological study 
of the model. So we do not discuss this problem further here. 
It should be noted that the existence of the different singlet 
fields $S_{1,2}$ are
generally required in order to make $\chi$ and $\bar\chi$ 
play a role of messengers of the
supersymmetry breaking. In fact, the $F^\prime$ charges of $\chi$ and
$\bar\chi$ satisfy
\begin{equation}
e-f=-5(b+d)\not=0,    \qquad ({\rm mod}~n)
\end{equation}
which is derived from eq.~(\ref{mess}). This feature is caused by the
direct product structure of the gauge group which is motivated 
to realize the doublet-triplet splitting.  
 
We can now consider the physics at the scale after the symmetry breaking 
due to the VEVs in eqs.~(\ref{eqa}) and (\ref{eqa0}). The massless degrees of
freedom are composed of the contents of the MSSM and the fields 
$(q, l)$ and $(\bar q, \bar \ell)$ which come from $\chi({\bf 1}, {\bf 5})$
and $\bar\chi(\bar{\bf 5}, {\bf 1})$.
We can expect the successful gauge coupling unification for these 
field contents in the similar way to the MSSM.
Under the discrete symmetry $F^\prime$, the superpotential for these
fields can be written as
\begin{equation}
W_2=h_1\Psi_{10}\Psi_{10}H_2 + 
h_2\Psi_{10}\Psi_{\bar 5}\tilde H_{2} 
+\lambda_1 S_1q\bar q +\lambda_2 S_2\ell\bar\ell. 
\end{equation}
The last three terms effectively appear through the nonrenormalizable terms 
as a result of the symmetry breaking due to $\langle\phi_1\rangle$
and $\langle\phi_2\rangle$. 
This feature makes the second term favorable to explain the hierarchy 
between the masses of the top and bottom quarks \cite{w}.
The messenger fields $q,~\bar q$ and $\ell,~\bar\ell$ couple with 
the different singlet fields $S_{1,2}$.
If both their scalar components and $F$-components get the VEVs,
they can play the role of messenger fields for the supersymmetry 
breaking in the observable sector as in the minimal gauge mediation
model \cite{mgm1,mgm2}. This is discussed in the next section. 
Although it seems to be difficult to produce a weak scale $\mu$-term 
within the field contents given in Table 1, 
it may be expected to be generated associated with 
the supersymmetry breaking by extending the model.
We may consider various ways for generating the $\mu$-term such as 
the mechanism of Giudice-Masiero \cite{gm} or the model based 
on the VEV of the singlet field like the next MSSM \cite{mu,mgm2}. 
However, we do not discuss its origin here and treat it only as 
an effective parameter in the following discussion. 

\section{Soft SUSY breaking parameters}
In this section we study the soft supersymmetry breaking parameters 
in the present model.
The gauge anomaly cancellation for ${\cal G}$ requires us to introduce 
a set of vectorlike fields $\chi$ and $\bar\chi$ as mentioned above.
Using these fields as the messenger fields, fortunately, we can 
apply the well-known minimal gauge mediation supersymmetry breaking 
scenario \cite{mgm1,mgm2,exmgm1,exmgm2} to our model.
If the singlet chiral superfields $S_1$ and $S_2$ couple with the 
hidden sector fields which break down the supersymmetry,
$q, \bar q$ and $\ell, \bar\ell$ play the role of messenger fields as
in the ordinary scenario.
The only difference from the ordinary minimal gauge mediation scenario is that 
$q,~\bar q$ and $\ell,~\bar\ell$ couple with different singlet chiral 
superfields $S_1$ and $S_2$ in the superpotential $W_2$
because of the discrete symmetry $F^\prime$. 
If we assume that both $S_\alpha$ and $F_{S_\alpha}$ 
get the VEVs, the gaugino masses and the soft scalar masses are 
generated through one-loop and two-loop diagrams, respectively. 
However, the mass formulas are modified from the usual ones
since the messenger fields $q,~\bar q$ and $\ell,~\bar\ell$ 
couple with the different singlets.
\input epsf 
\begin{figure}[tb]
\begin{center}
\epsfxsize=11.4cm
\leavevmode
\epsfbox{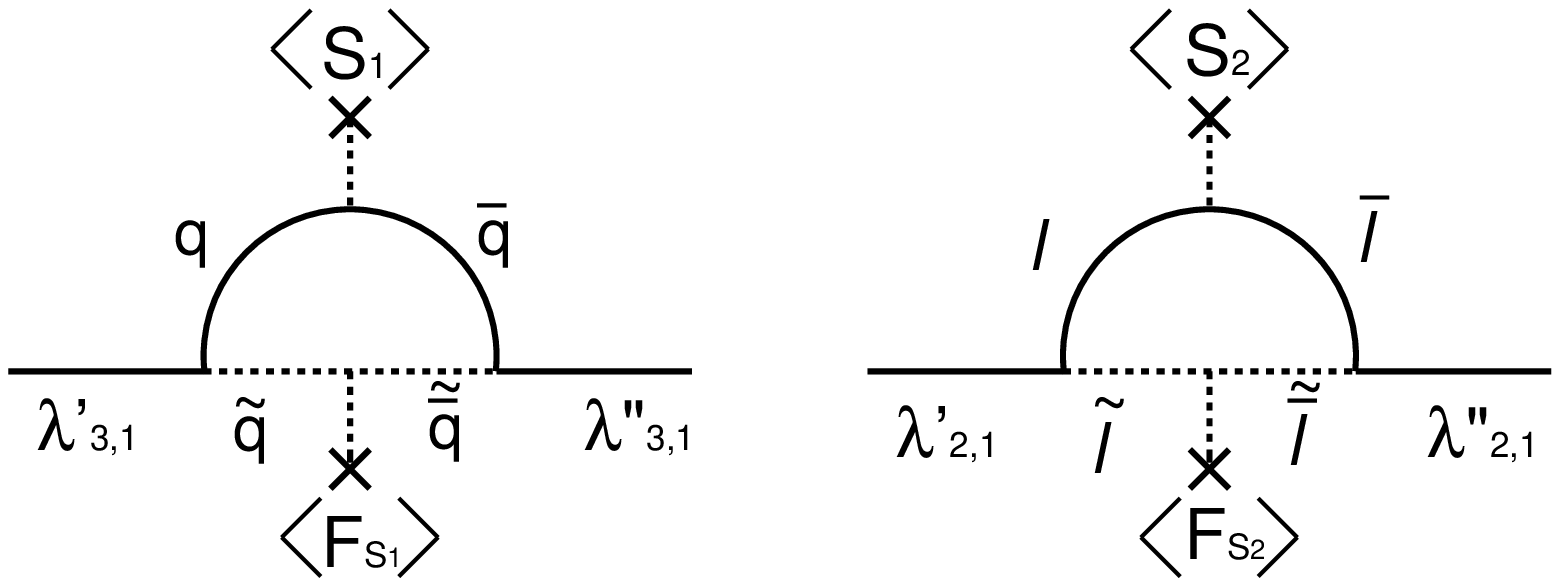}\\
{\footnotesize Fig. 1~~\  One loop diagrams contributing to 
gaugino masses.}
\end{center}
\end{figure}

The massless vector supermultiplet of SU(5) is written as
\begin{equation}
V= V^\prime\cos\theta + V^{\prime\prime}\sin\theta,
\label{eqb}
\end{equation}
where $V^\prime$ and $V^{\prime\prime}$ are the vector supermultiplets of
SU(5)$^\prime$ and SU(5)$^{\prime\prime}$. A mixing angle $\theta$ and a new
gauge coupling constant $g$ of SU(5) are determined as
\begin{equation}  
{1\over g^2}={1\over g^{\prime 2}}+{1\over g^{\prime\prime 2}}, \qquad
\tan\theta={g^\prime\over g^{\prime\prime}},
\label{eqc}
\end{equation}
where $g^\prime$ and $g^{\prime\prime}$ are the gauge coupling constants 
of SU(5)$^\prime$ and SU(5)$^{\prime\prime}$.
The same relations are satisfied for each factor group of ${\cal H}$
at the symmetry breaking scale $\tilde M$.
The gauge coupling constants of ${\cal H}$ follow the 
unification relation $g=g_3=g_2=\sqrt{5\over 3}g_1$ at $\tilde M$.
The information on the direct product gauge structure at the 
high energy region 
is included in the mixing angle $\theta$.
The gauginos become massive due to the mixing between the gauginos of 
SU(5)$^\prime$ and SU(5)$^{\prime\prime}$ through the one-loop diagrams
shown in Fig.~1. 
These mass mixings can be estimated as
\begin{equation}
M_{\lambda_3^\prime\lambda_3^{\prime\prime}}
={g_3^\prime g_3^{\prime\prime}\over 16\pi^2}\Lambda_1, \qquad
M_{\lambda_2^\prime\lambda_2^{\prime\prime}}
={g_2^\prime g_2^{\prime\prime}\over 16\pi^2}\Lambda_2, \qquad
M_{\lambda_1^\prime\lambda_1^{\prime\prime}}
={g_1^\prime g_1^{\prime\prime}\over 16\pi^2}
\left({2\over 3}\Lambda_1+\Lambda_2\right),
\label{eqf}
\end{equation}
where $\Lambda_\alpha=\langle F_{S_\alpha}\rangle/\langle S_\alpha\rangle$.
These can be transformed into the masses $M_r$ of the gauginos 
$\lambda_r$ of the gauge group ${\cal H}$ by taking account of 
eqs.~(\ref{eqb}) and (\ref{eqc}). 
They can be written in the form as
\begin{equation}
M_3={\alpha_3\over 4\pi}\Lambda_1, \qquad
M_2={\alpha_2\over 4\pi}\Lambda_2, \qquad
M_1={\alpha_1\over 4\pi}\left({2\over 3}\Lambda_1+\Lambda_2\right), 
\label{eqff}
\end{equation}
where $\alpha_r=g_r^2/4\pi$.

These give the same sum rule among gaugino masses as
the one of the usual minimal gauge mediation scenario at the supersymmetry
breaking scale such as
\begin{equation}
{M_1\over \alpha_1}={2\over 3}{M_3\over \alpha_3} +{M_2\over \alpha_2}.
\end{equation}
However, depending on the ratio of the scale $\Lambda_1/\Lambda_2$,
each mass ratio can be different from the ordinary ones, that is,
\begin{equation}
{M_2\over M_3}={\alpha_2\over \alpha_3}{\Lambda_2\over\Lambda_1}, \qquad
{M_1\over M_3}={\alpha_1\over \alpha_3}\left({2\over 3}
+{\Lambda_2\over\Lambda_1}\right).
\end{equation} 
These formulas show that $M_3$ can be much smaller than $M_{1,2}$ in the
case of $\Lambda_2>\Lambda_1$. 
If we take account of the evolution effect by the renormalization 
group, their values at the weak scale $M_W$, for example, can be obtained as
\begin{equation}
M_r(M_W)=M_r(\Lambda){\alpha_r(M_W)\over\alpha_r(\Lambda)},
\end{equation}
where $\Lambda$ is a scale at which the supersymmetry breaking is introduced.
Since $\Lambda_\alpha$ is generally independent, the phases contained in the 
gaugino masses are non-universal even in the case of 
$\vert\Lambda_1\vert=\vert\Lambda_2\vert$. We cannot remove them completely 
by using the $R$-transformation unlike the universal gaugino mass case.
In fact, if we define the phases as
$\Lambda_\alpha\equiv\vert\Lambda_\alpha\vert e^{i\varphi_\alpha}$ and make
$M_2$ real by the $R$-transformation, the phases of $M_3$ and $M_1$ are
written as
\begin{equation}
{\rm arg}(M_3)=\varphi_1-\varphi_2, \qquad 
{\rm arg}(M_1)=\arctan\left({2\vert\Lambda_1\vert
\sin(\varphi_1-\varphi_2)\over
3\vert\Lambda_2\vert+2\vert\Lambda_1\vert
\cos(\varphi_1-\varphi_2)}\right).
\end{equation}

The scalar masses are induced as the values of
$O(\vert\Lambda_\alpha\vert^2)$ through the two-loop diagrams as in 
the ordinary case. Again, only difference comes from the fact that the
model has the direct product gauge structure at the high energy region
and the messengers $(q,~\ell)$ and $(\bar q,~\bar\ell)$ 
are the representations of the different factor groups.
This brings the mixing factor between the vector superfields $V_r$ and
$V_r^\prime$, $V_r^{\prime\prime}$ as in the gaugino mass case.
Taking account of this, their formulas can be written as
\begin{equation}
\tilde m^2_f=2\left[C_3\left({\alpha_3\over 4\pi}\right)^2 
+{2\over 3}\left({Y\over 2}\right)^2\left({\alpha_1\over 4\pi}\right)^2\right]
\vert\Lambda_1\vert^2 
+2\left[C_2\left({\alpha_2\over 4\pi}\right)^2 
+\left({Y\over 2}\right)^2\left({\alpha_1\over 4\pi}\right)^2\right]
\vert\Lambda_2\vert^2,
\label{eqg}
\end{equation}
where $C_3=4/3$ and 0 for the SU(3) triplet and singlet fields, and
$C_2=3/4$ and 0 for the SU(2) doublet and singlet fields, respectively. 
The hypercharge $Y$ is expressed as $Y=2(Q-T_3)$. 
These formulas can give rather different mass spectrum for the scalar
fields depending on the values of $\Lambda_1/\Lambda_2$. In fact,
if we assume $\Lambda_1< \Lambda_2$, for example,
the mass difference between the color singlet fields and 
the colored fields tend to be smaller in comparison with the one in the
ordinary scenario.
Let us take $\Lambda_1=40$~TeV and $\Lambda_2=100$~TeV to
show a typical spectrum of the superpartners at the supersymmetry
breaking scale. Then we can have the following spectrum as
\begin{eqnarray}
&& M_3=273~{\rm GeV}, \quad   M_2=279~{\rm GeV}, \quad M_1=111~{\rm GeV}, \quad
\tilde m_Q=562~{\rm GeV}, \nonumber \\
&& \tilde m_U=455~{\rm GeV}, \quad \tilde m_D=449~{\rm GeV}, \quad  
\tilde m_L=347~{\rm GeV}, \quad \tilde m_E=130~{\rm GeV}, \nonumber \\ 
&&m_1=m_2=347~{\rm GeV},  
\label{eqh}
\end{eqnarray}
where $m_1$ and $m_2$ are masses of the Higgs scalars that couple 
with the down and up sectors of quarks and leptons, respectively.
These masses are somewhat affected by the renormalization group running
effect, although the running region is not so large. 
For example, the modifications due to this effect can be solved 
analytically for the masses of sleptons and $H_1$, for which Yukawa 
coupling effects can be neglected, as \cite{exmgm2}
\begin{eqnarray}
&&\tilde m_{\ell_L}^2(M_W)=\tilde m_{\ell_L}^2(\Lambda)
-{3\over 2}\vert M_2(\Lambda)\vert^2
\left({\alpha^2_2(M_W)\over\alpha_2^2(\Lambda)}-1\right)
-{1\over 22}\vert M_1(\Lambda)\vert^2
\left({\alpha^2_1(M_W)\over\alpha_1^2(\Lambda)}-1\right), \nonumber \\
&&\tilde m_{\ell_R}^2(M_W)=\tilde m_{\ell_R}^2(\Lambda)
-{2\over 11}\vert M_1(\Lambda)\vert^2
\left({\alpha^2_1(M_W)\over\alpha_1^2(\Lambda)}-1\right), 
\label{eqi}
\end{eqnarray}
where we do not write the $D$-term contribution explicitly.
The mass $m_1^2$ of the Higgs scalar has the same formula as 
$\tilde m_{\ell_L}^2$.

As in the minimal gauge mediation model discussed in \cite{exmgm2},
the soft supersymmetry breaking $A_f$ and $B$ parameters can also be expected
to be induced through the radiative correction such as
\begin{eqnarray}
&&A_f\simeq A_f(\Lambda)+M_2(\Lambda)\left(-1.85+0.34\vert h_t\vert^2\right),
\nonumber\\
&&{B\over\mu}\simeq {B\over\mu}(\Lambda)-{1\over 2}A_t(\Lambda)
+M_2(\Lambda)\left(-0.12+0.17\vert h_t\vert^2\right),
\label{eqj}
\end{eqnarray}
where we should omit a term of $h_t$ in the expression of 
$A_f$ except for the top sector $(f=t)$. 
In the case of $A_f(\Lambda)=B(\Lambda)=0$ which are expected in many
gauge mediation scenario, $A_f$ and $B$ are proportional to 
$M_2$ and then the $CP$ phases in the soft supersymmetry breaking
parameters are completely rotated away as far as gaugino masses are 
universal \cite{exmgm2}.
However, in the present model this situation is broken even in the case of 
$A_f(\Lambda)=B(\Lambda)=0$ since the phases in the gaugino masses are not 
universal. 
Although the generation of $B$ should be considered on the basis 
of the various mechanisms like the $\mu$-term \cite{mu} also 
in the present model, it is completely 
model dependent and we do not discuss it further. 

Finally we comment on some phenomenological aspects on these soft breaking
parameters.
At present it seems to be difficult to relate the supersymmetry breaking
parameters to the observed values. Only exception might be found 
in the electroweak symmetry breaking. As is well-known ,  
the minimum condition
of the tree level scalar potential in the MSSM can be written as
\begin{equation}
m_Z^2=-2\mu^2+2~{m_1^2-m_2^2\tan^2\beta\over \tan^2\beta -1}.
\label{eqk}
\end{equation}
Supersymmetry breaking parameters in the right-hand side can be
estimated by using the one-loop renormalization 
group equations (RGEs). Through the semi-analytic calculation \cite{ccopw},
their weak scale values can be expressed using various soft parameters at
the supersymmetry breaking scale $\Lambda$ whose examples are shown in 
eq.~(\ref{eqi}). If we take $\Lambda=100$~TeV, $m_{\rm top}=170$~GeV and 
$\tan\beta=5$ and use the numerical coefficient obtained through the RGEs
in this case \cite{klnw}, 
eq.~(\ref{eqk}) is expressed as\footnote{In this expression we do not
take account of the difference between $\Lambda_1$ and $\Lambda_2$
for the estimation of the numerical coefficients.
However, we can expect that there is no substantial difference even if we
take account of it.}
\begin{equation}
m_Z^2=-1.8\mu^2-0.2M_2^2+0.4M_3^2+0.2A_t^2+0.4\tilde m_{Q_3}^2
+0.4\tilde m_{U_3}^2-1.7m_2^2-0.2A_tM_3+\cdots,
\end{equation}
where the ellipses represent the subdominant contributions.
Assuming $\Lambda_{1,2}$ to be real and
substituting soft parameters given by eqs.(\ref{eqff}) and (\ref{eqg}),
we obtain
\begin{equation}
m_Z^2=(115x^2+6.1x-13.2)(10^{-3}\Lambda_2)^2-1.8\mu^2,
\end{equation} 
where $x=\Lambda_1/\Lambda_2$ and we use eq.~(\ref{eqj}) 
with $A_t(\Lambda)=0$. In Fig.2 we plot the values of $\mu$ satisfying
this relation for various values of $\Lambda_{1,2}$.
This shows that $\mu$ can take a reasonable value as far as we can set up
$\Lambda_{1,2}$ appropriately. As a general feature we find that
the large $\Lambda_2$ tend to require the large value of $\mu$. 
The sensitivity of $\mu$ against $x$ seems to be almost independent 
of the value of $x$ in the $x~{^>_\sim}~0.5$ region.
In the $x~{^<_\sim}~0.5$ region, we can obtain a small value of
$\mu$ such as $\mu\sim 100~GeV$ as the consistent solution.
However, it is necessary to tune carefully the value of $x$ 
to be $0.35\sim 0.5$ depending on $\Lambda_2$
to obtain the smaller value of $\mu$.
This required tuning is finer for the larger $\Lambda_2$ value.  
Anyway, this feature looks different from
the ordinary minimal gauge mediation model in which $x=1$ is satisfied
and then $\mu~{^>_\sim}~300$~GeV is required for $\Lambda_2\ge 40$ TeV 
as seen from Fig.2.\footnote{The small $\Lambda_2$ makes $M_2$ too small
and it will be excluded from the fact that neutralinos and charginos 
have not been found at LEP.}   
\begin{figure}[tb]
\begin{center}
\epsfxsize=8.4cm
\leavevmode
\epsfbox{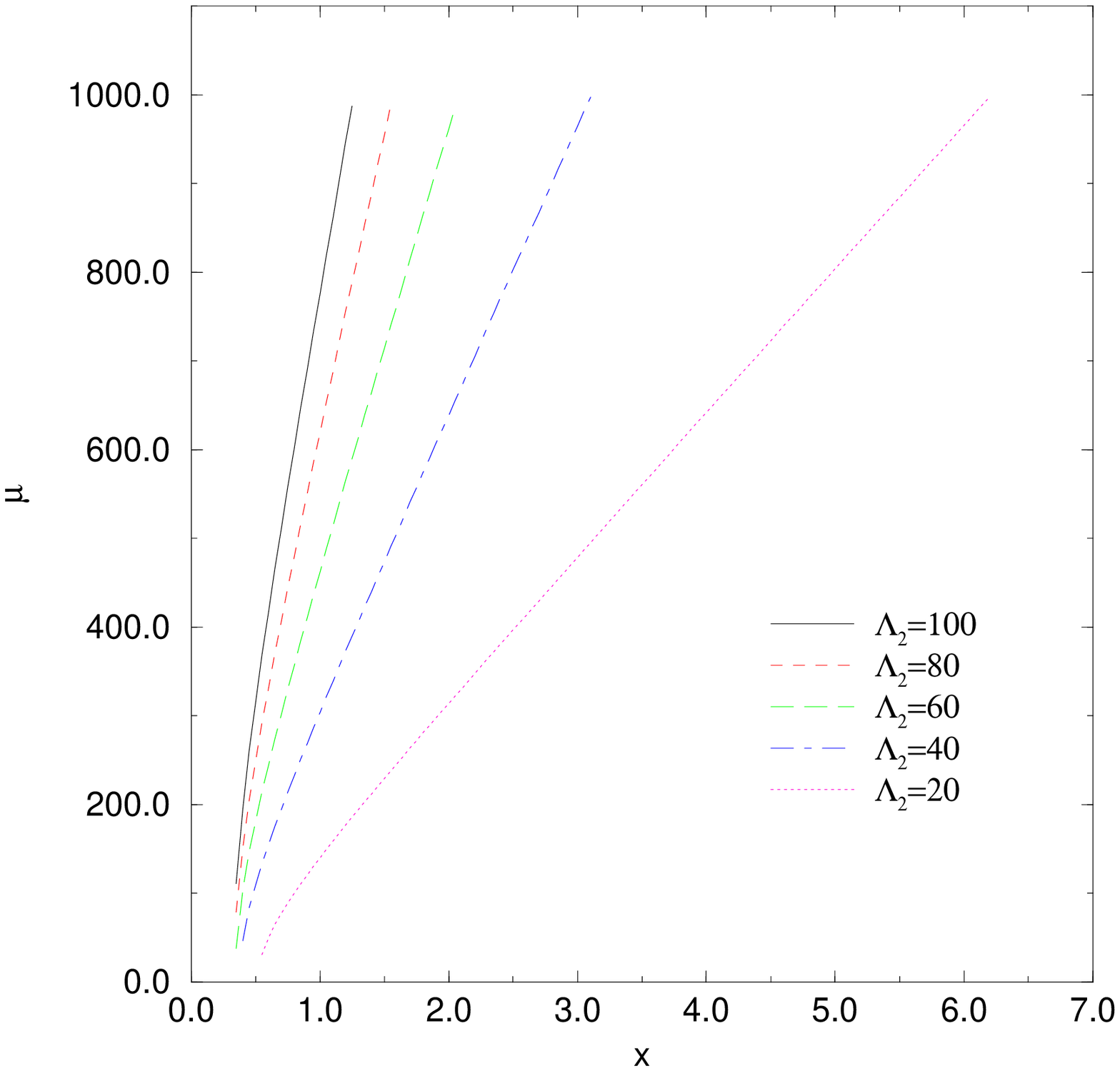}\\
\vspace*{-0.6cm}
{\footnotesize Fig. 2~~\  Values of $\mu$ required to realize a
 correct vacuum for the various SUSY breaking scales $\Lambda_{1,2}$~TeV.}
\end{center}
\end{figure}

There is another interesting feature in this case.
In the usual minimal gauge mediation scenario the lightest superparticle 
in the whole spectrum except for the gravitino is the right-handed
slepton as far as we do not take account of the radiative effect.
However, in the present scenario the photino can be the lightest one
in this situation even at the tree level as we can see it in the example 
given in eq.(\ref{eqh}).
This feature might be relevant to the event such as $e^+e^-\gamma\gamma+$
missing $E_T$ \cite{exmgm1,exmgm2}. 
Our model might be discriminated from other gauge mediation models by 
using this aspect. 

The gaugino mass universality seems to be a rather general result in
various supersymmetry breaking scenarios. However, the present model 
naturally induces non-universal gaugino masses as a result of intrinsic
nature of the model. We generally have physical $CP$ phases in the gaugino
sector. This may be dangerous since it can give a
large contribution to the electric dipole moment of a neutron and
an electron. However, they could be within the experimental bound 
even if the $CP$ phases are $O(1)$. It is expected that 
there can be an effective cancellation between the chargino and 
neutralino contributions to them \cite{edm}.
We can check this in the present model and the result will be presented
elsewhere \cite{ts}. In the case that there is no contradiction with the 
EDM, these large $CP$ phases may be important when we
consider the electroweak baryogenesis \cite{baryo}.    
 
\section{Summary}
We investigated the soft supersymmetry breaking masses in the
supersymmetric unified model which can solve the doublet-triplet 
splitting problem.
The model is constructed through the deconstruction by extending the
gauge structure into the direct product group 
SU(5)$^\prime\times$SU(5)$^{\prime\prime}$. 
The low energy spectrum is the one of the MSSM with the 
additional chiral superfields which can play a role of messengers 
in the gauge mediated supersymmetry breaking.
The gauge anomaly cancellation requires to introduce these 
chiral superfields. The discrete symmetry can be introduced to realize the 
doublet-triplet splitting because of the the direct product gauge structure.
It forces the color triplet and color singlet
messengers to couple with the different singlet chiral superfields whose 
scalar and auxiliary components are assumed to get the VEVs due to 
the hidden sector dynamics.
This can make the different structure of the soft supersymmetry breaking 
masses from the ones of the ordinary minimal gauge mediation
scenario. One of the interesting feature is that the mass difference
between the colored fields and the color singlet fields can be smaller 
in comparison with the ordinary gauge mediation scenario.
Another interesting point is that the gaugino masses become
non-universal generally and the non-universal phases are introduced 
in the gaugino masses. 
The $CP$ phases can remain in the gaugino sector as the physical phases 
after the $R$ transformation. This feature may discriminate this model
from others since it is rather difficult to construct the well-motivated 
model with the non-universal gaugino masses.
Further phenomenological study of the model seems to be worthy since
it is constructed on the basis of the reasonable motivation to solve 
the doublet-triplet splitting problem in the grand unified model.

\vspace{.5cm}
\noindent
This work is supported in part by a Grant-in-Aid for Scientific 
Research (C) from Japan Society for Promotion of Science
(No.~14540251) and also by a Grant-in-Aid for Scientific 
Research on Priority Areas (A) from The Ministry of Education, Science,
Sports and Culture (No.~14039205).


\end{document}